\documentclass[iop,apj,tighten]{emulateapj}

\shorttitle{INTERSTELLAR SODIUM AND CALCIUM TOWARD SN 2011DH}
\shortauthors{RITCHEY \& WALLERSTEIN}

\begin{document}
\title{Interstellar Sodium and Calcium Absorption toward SN 2011dh in M51}
\author{Adam M. Ritchey\altaffilmark{1} and George Wallerstein\altaffilmark{1}}
\altaffiltext{1}{Department of Astronomy, University of Washington, Seattle, 
WA 98195, USA; aritchey@astro.washington.edu; wall@astro.washington.edu}

\begin{abstract}
We present high-resolution echelle observations of SN~2011dh, which exploded 
in the nearby, nearly face-on spiral galaxy M51. Our data, acquired on three 
nights when the supernova was near maximum brightness, reveal multiple 
absorption components in Na~{\sc i}~D and Ca~{\sc ii}~H and K, which we 
identify with gaseous material in the Galactic disk or low halo and in the 
disk and halo of M51. The M51 components span a velocity range of over 
140~km~s$^{-1}$, extending well beyond the range exhibited by H~{\sc i} 21~cm 
emission at the position of the supernova. Since none of the prominent 
Na~{\sc i} or Ca~{\sc ii} components appear to coincide with the peak in 
H~{\sc i} emission, the supernova may lie just in front of the bulk of the 
H~{\sc i} disk. The Na~{\sc i}/Ca~{\sc ii} ratios for the components with the 
most extreme positive and negative velocities relative to the disk are 
$\sim$1.0, similar to those for more quiescent components, suggesting that the 
absorption originates in relatively cool gas. Production scenarios involving a 
galactic fountain and/or tidal interactions between M51 and its companion 
would be consistent with these results. The overall weakness of Na~{\sc i}~D 
absorption  in the direction of SN~2011dh confirms a low foreground and host 
galaxy extinction for the supernova.
\end{abstract}


\section{INTRODUCTION}
Bright supernovae occurring in external galaxies afford a unique opportunity 
to probe the structure of interstellar gas in the Galactic halo as well as in 
the supernova host galaxy through high-resolution, absorption-line 
spectroscopy. In recent decades, a number of supernovae have been used in this 
manner, providing insights on the characteristics of gaseous material in 
nearby galaxies, in the intergalactic medium, and in Galactic high-velocity 
clouds, or HVCs (e.g., Jenkins et al. 1984; D'Odorico et al. 1989; Steidel et 
al. 1990; Meyer \& Roth 1991; Bowen et al. 1994; Vladilo et al. 1994; Ho \& 
Filippenko 1995; King et al. 1995; Bowen et al. 2000).

The recent discovery of the Type IIb SN~2011dh in M51 (Arcavi et al. 2011; see 
also Griga et al. 2011; Silverman et al. 2011), which reached a maximum 
brightness of $V\approx12.5$, presented us with the chance to acquire 
high-resolution spectra to search for interstellar absorption lines in the 
direction of the supernova. This particular supernova is significant in that 
it is the third such object to be detected in M51 (the Whirlpool Galaxy; 
NGC~5194) within the past 20 years, following the discoveries of SN~2005cs (a 
Type IIP) and SN~1994I (a Type Ic). Ho \& Filippenko (1995) obtained 
high-resolution observations of SN~1994I, finding numerous strong absorption 
components of Na~{\sc i}~D associated with interstellar gas in M51, along with 
components of more moderate strength, some identified with the disk of our 
Galaxy and others likely identified as HVCs belonging to either our Galaxy or 
M51.

In this Letter, we report on high-resolution observations of Na~{\sc i}~D and 
Ca~{\sc ii}~H and K absorption along the line of sight to SN~2011dh, which 
exploded at an apparent position $2^{\prime}.1$ E and $1^{\prime}.5$ S of the 
nucleus of M51 (Van Dyk et al. 2011). This corresponds to a projected distance 
of 6.1~kpc, assuming the distance to M51 is 8.0~Mpc (as listed in the 
NASA/IPAC Extragalactic Database\footnote{http://ned.ipac.caltech.edu/}). The 
positions of SN~2011dh and SN~1994I (which was located $0^{\prime}.3$ or 
0.7~kpc from the nucleus) are separated by only $2^{\prime}.3$ on the sky. 
Thus, by comparing our results with those of Ho \& Filippenko (1995), we are 
able to simultaneously probe small-scale structure in the Galactic disk and 
halo and galactic-scale structure in the interstellar medium (ISM) of M51.

\section{OBSERVATIONS AND DATA REDUCTION}
Our data were acquired with the Astrophysical Research Consortium (ARC) 3.5~m 
telescope at Apache Point Observatory (APO), using the echelle spectrograph, 
which provides complete wavelength coverage in the range 3800--10200 \AA{} 
with a resolving power of $R\approx31,500$ ($\Delta v\sim9.5$ km s$^{-1}$). 
Three 30 minute exposures of SN~2011dh were obtained on 2011 June 9, one on 
2011 June 13, and three more on 2011 June 20 (all dates are given in UT), 
during which time the supernova increased from $V\approx13.2$ to 
$V\approx12.6$. Evidently, our observations on June 20 were made shortly 
before maximum brightness. The bright, nearby star $\eta$~UMa (B3 V; $V=1.85$) 
also was observed on June 9 and June 20 to aid in the removal of telluric 
absorption lines in the spectrum of SN~2011dh.

\begin{figure}
\centering
\includegraphics[width=0.45\textwidth]{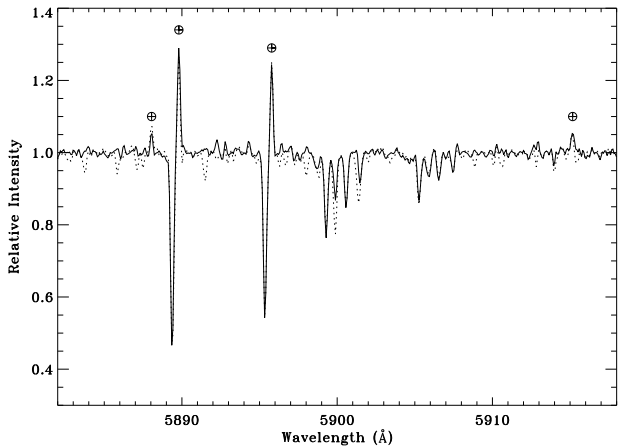}
\caption[]{Coadded spectra of SN 2011dh in the vicinity of Na~{\sc i}~D. The 
final spectrum obtained after removing telluric absorption lines from 
individual exposures is indicated by the solid line. The dotted line shows how 
the spectrum would appear if no telluric lines were removed. Note that some of 
the telluric features affect the Na~{\sc i}~D components near 5900~\AA{} (the 
D$_2$ lines associated with M51; see Figure~2), but that the removal procedure 
yields the appropriate relative strengths of these lines compared to the 
weaker lines of the doublet. Strong night sky emission lines are identified 
with the symbol $\oplus$. These were removed before proceeding with profile 
fitting.}
\end{figure}

The raw exposures were reduced using a semi-automated pipeline reduction 
procedure, which employs standard IRAF routines for bias correction, 
cosmic-ray removal, scattered-light subtraction, one-dimensional spectral 
extraction, flat-fielding, and wavelength calibration. The reduction procedure 
is based on the IRAF Data Reduction Guide for the ARC Echelle Spectrograph 
written by J. 
Thorburn\footnote{http://www.apo.nmsu.edu/arc35m/Instruments/ARCES/}. 
Atmospheric absorption lines were removed from individual, calibrated 
exposures of SN 2011dh with the IRAF task {\sc telluric} using $\eta$~UMa as 
the telluric standard. This procedure accounts for differences in airmass as 
well as in the abundances of telluric species between the standard spectrum 
and the exposure being corrected (see Figure~1 for a demonstration of the 
effectiveness of the telluric line removal procedure in the vicinity of 
Na~{\sc i}~D). The corrected spectra were then shifted to the reference frame 
of the local standard of rest (LSR) and the individual exposures were co-added 
to produce a final, high signal-to-noise ratio (S/N) spectrum of SN~2011dh. 
The S/N ranges from $\sim$110 near 5900~\AA{} to $\sim$40 near 3900~\AA.

The only interstellar absorption features positively identified in our final 
spectrum of SN~2011dh are those of the Na~{\sc i}~D (D$_1$~$\lambda$5895; 
D$_2$~$\lambda$5889) and Ca~{\sc ii}~H ($\lambda$3968) and K ($\lambda$3933) 
doublets. The overlapping echelle orders containing these lines were combined 
to improve the S/N and the resulting spectra were normalized to the relatively 
featureless continuum. As seen in Figure~2, we detect Na~{\sc i} and 
Ca~{\sc ii} absorption both from the Milky Way (near 
$v_{\rm LSR}=0$~km~s$^{-1}$) and from M51 (near the M51 systemic velocity of 
$v_{\rm LSR}=468$~km~s$^{-1}$; Walter et al. 2008, applying a 
heliocentric-to-LSR velocity correction of 12~km~s$^{-1}$). Given the 
overall weakness of these features, it is not surprising that other optical 
absorption lines, such as K~{\sc i}~$\lambda7698$, Ca~{\sc i}~$\lambda4226$, 
CH$^+$~$\lambda4232$, and CH~$\lambda4300$, are not detected. However, we are 
unable to search for K~{\sc i}~$\lambda7698$ absorption at the redshift of M51 
due to the presence of strong night sky emission lines at the same wavelengths.

Before analyzing the final combined spectrum of SN 2011dh, we compared the 
nightly coadded spectra from June 9 and June 20 --- the two nights on which we 
have the best S/N due to our having obtained multiple exposures --- in order 
to search for variations in the absorption profiles. A variable component 
might be expected if the absorption originates in circumstellar material 
associated with the progenitor of SN 2011dh, which gets swept up in the 
supernova blast wave. Upon careful inspection, we find no evidence for 
significant variation in any of the Na~{\sc i} or Ca~{\sc ii} components. 
Thus, the results of this paper are based solely on the analysis of the 
combined spectrum.

\section{ANALYSIS}
The interstellar Na~{\sc i} and Ca~{\sc ii} absorption profiles toward 
SN~2011dh were analyzed by means of the profile fitting routine ISMOD 
(Y. Sheffer, unpublished), which fits multiple Voigt components to the 
observed spectrum, convolving the intrinsic line profile with an instrumental 
profile, assumed to be Gaussian in shape. Using a simple rms-minimizing 
technique, ISMOD determines best-fit values for the velocity, $b$-value, and 
column density of each component. Since the absorption features are relatively 
weak, line saturation is not a major concern for these data. Only the 
Na~{\sc i} component near $v_{\rm LSR}=-30$ km s$^{-1}$, which presumably 
arises from gas within the Milky Way, shows significant optical depth at line 
center. For this component, the ratio of the Na~{\sc i}~D$_2$ to D$_1$ 
equivalent widths is 1.2 (as opposed to 2.0 in the optically thin case) and 
implies a Doppler width of $b=2.1\pm0.2$~km~s$^{-1}$. In order to derive an 
accurate column density for this mildly-saturated component, the $b$-value was 
held fixed at 2.1~km~s$^{-1}$ in the profile fit for both the D$_2$ and D$_1$ 
lines.

Our analysis of Ca~{\sc ii} absorption toward SN 2011dh rests mostly on the 
Ca~{\sc ii}~K line, because the weaker H components are strongly affected by 
noise in the spectrum. We initially fit the Na~{\sc i}~D and Ca~{\sc ii}~K 
profiles by focusing on the prominent components seen in both species. 
Ultimately, we were required to include additional components to account for 
the extra Ca~{\sc ii} absorption surrounding the main Galactic component and 
to fill in gaps between the stronger Na~{\sc i} and Ca~{\sc ii} components 
associated with M51. To aid in the process of decomposing the blended 
profiles, the $b$-values were constrained to fall within certain limits 
consistent with ultra-high resolution surveys of the Galactic ISM (e.g., 
Welty et al. 1994, 1996). We note, however, that $b$-values derived in this 
study are more appropriately referred to as ``effective'' $b$-values since at 
our resolution ($\Delta~v\sim9.5$~km~s$^{-1}$) unresolved substructure almost 
certainly exists in the data. The results of fitting the Ca~{\sc ii}~K line 
were then applied to Ca~{\sc ii}~H to check for consistency, holding the 
relative velocities, $b$-values, and fractional column densities fixed.

\begin{figure*}
\centering
\includegraphics[width=0.9\textwidth]{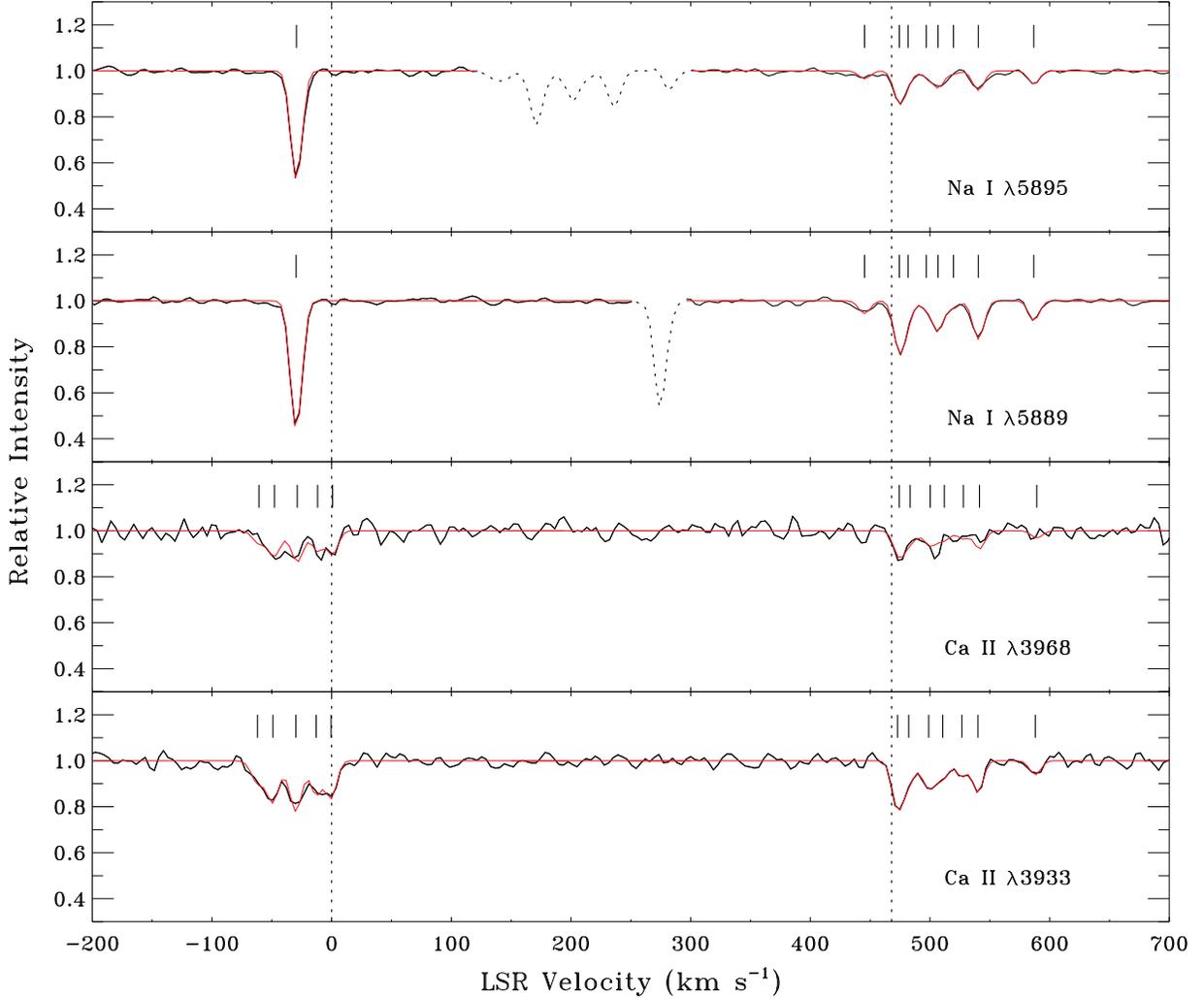}
\caption[]{Normalized high-resolution spectra of SN 2011dh in the vicinity of 
the Na~{\sc i}~D and Ca~{\sc ii}~H and K lines. Synthetic absorption profiles 
(red curves) are shown superimposed onto the observed spectra (black lines). 
Dotted features denote overlap in the Na~{\sc i}~D lines. Tick marks indicate 
the positions of the fitted components. Vertical dotted lines mark the local 
rest velocity of the Galaxy ($v_{\rm LSR}=0$~km~s$^{-1}$) and the systemic 
velocity of M51 ($v_{\rm LSR}=468$~km~s$^{-1}$).}
\end{figure*}

Table~1 presents the Na~{\sc i} and Ca~{\sc ii} equivalent widths (and 
1$\sigma$ uncertainties) derived from our profile fits to these lines (see 
Figure~2). In Table~2, we give the fitted column density and $b$-value of each 
Na~{\sc i}~D and Ca~{\sc ii}~K component along with the mean LSR velocity. In 
most cases, the uncertainties in $N$ reflect the uncertainties in 
$W_{\lambda}$, which are based on the rms deviations in the continuum and the 
widths (FWHM) of the absorption features. These uncertainties effectively 
include both photon noise and errors in continuum placement. For the 
Na~{\sc i} component near $v_{\rm LSR}=-30$ km s$^{-1}$, the column density 
uncertainties also include the effects of varying the $b$-value within the 
range allowed by the errors in the doublet ratio. Final Na~{\sc i} column 
densities were determined by taking the weighted mean of the column densities 
derived from the D$_2$ and D$_1$ lines, while for Ca~{\sc ii} we simply 
adopted the more precise results from Ca~{\sc ii}~K. The agreement with column 
densities from Ca~{\sc ii}~H is at the 10\% level or better. The last column 
of Table~2 presents the $N$(Na~{\sc i})/$N$(Ca~{\sc ii}) ratio (or limits on 
the ratio) for each of the Milky Way and M51 components.

We find total Na~{\sc i}~D$_2$ and D$_1$ equivalent widths of 
$180.1\pm5.0$~m\AA{} and $106.2\pm5.1$~m\AA{}, respectively, for the features 
associated with M51. These values are in good agreement with those reported by 
Arcavi et al. (2011), which were obtained from Keck High Resolution Echelle 
Spectrometer data. As those authors noted, such weak absorption in the 
Na~{\sc i}~D lines indicates very little host galaxy extinction in the 
direction of SN 2011dh, probably less than the Galactic foreground extinction, 
which corresponds to an interstellar reddening of $E(B-V)=0.035$ (Schlegel et 
al. 1998).

\section{DISCUSSION}
An exciting aspect of this investigation is our ability to compare 
high-resolution optical spectra of two supernovae from the same external 
galaxy, allowing us to examine structure in the Galactic disk and halo at 
small angular scales as well as in the ISM of M51 at scales of several 
kiloparsecs. A comparison between our Na~{\sc i}~D results for SN 2011dh and 
those of Ho \& Filippenko (1995) for SN 1994I reveals both similarities and 
differences between the two lines of sight, which are separated by only 
$2^{\prime}.3$ on the sky (near Galactic coordinates $l=105^{\circ}$, 
$b=69^{\circ}$).

Both investigations find a strong Galactic Na~{\sc i} component at 
$v_{\rm LSR}\approx-30$ km s$^{-1}$ (system 1 in the nomenclature of Ho \& 
Filippenko 1995), which likely originates from relatively nearby gas, since 
there is very little difference in the Na~{\sc i} column density between the 
two sight lines. The $N$(Na~{\sc i})/$N$(Ca~{\sc ii}) ratio for this component 
($5.7\pm1.0$) is indicative of a high degree of Ca depletion, a characteristic 
of cool diffuse clouds at relatively low velocities (Siluk \& Silk 1974; Hobbs 
1983). Indeed, the LSR velocity of the component is consistent with the 
expected velocity of gas participating in Galactic rotation at $l=105^{\circ}$ 
(e.g., Sembach \& Danks 1994). We do not detect the weaker Galactic component 
at $v_{\rm LSR}\approx0$ km s$^{-1}$ (system 2 in Ho \& Filippenko) in 
Na~{\sc i}, but do observe this component, and other Galactic components at 
more negative velocities, in Ca~{\sc ii}. These additional Galactic components 
have only upper limits on $N$(Na~{\sc i})/$N$(Ca~{\sc ii}) --- suggesting low 
ratios ($<0.1$) --- and so presumably trace warmer gas, where Ca depletion is 
not as severe or where much of the neutral Na has been destroyed through 
collisional ionization.

Along with absorption components identified as belonging to the Milky Way, we 
find several components in Na~{\sc i} and Ca~{\sc ii} toward SN 2011dh with 
$v_{\rm LSR}\approx450$--$590$ km~s$^{-1}$. These are almost certainly 
associated with interstellar gas in M51, as are the much stronger Na~{\sc i} 
components that Ho \& Filippenko (1995) detect in the spectrum of SN 1994I 
with $v_{\rm LSR}\approx380$--$510$ km~s$^{-1}$ (their systems 6--14). Since 
SN 1994I occurred near the nucleus of M51, while SN 2011dh exploded in one of 
the outer spiral arms (see Figure~1 in Arcavi et al. 2011), the large 
difference in total Na~{\sc i} column density between the portions of M51 
probed by the two lines of sight is understandable. Like many spiral galaxies, 
the inner region of M51 exhibits only weak H~{\sc i} 21~cm emission (e.g., 
Rots et al. 1990). However, maps of CO 2$-$1 and 1$-$0 emission 
(Rand \& Kulkarni 1990; Schuster et al. 2007) indicate that the nucleus is 
rich in molecular gas, accounting for the higher Na~{\sc i} column density in 
that region.

\begin{figure}
\centering
\includegraphics[width=0.45\textwidth]{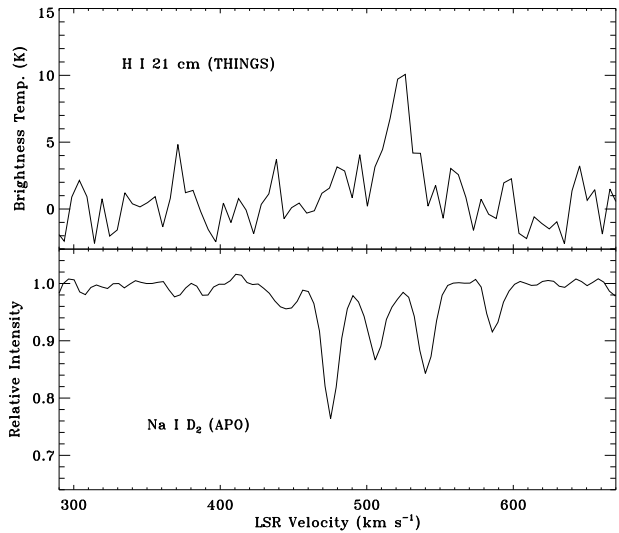}
\caption[]{Comparison between the H~{\sc i}~21~cm emission profile of M51 at 
the position of SN~2011dh provided by THINGS (upper panel) and the 
Na~{\sc i}~D$_2$ absorption profile from our APO observations (lower panel).}
\end{figure}

The strongest M51 component in our Na~{\sc i} and Ca~{\sc ii} data for 
SN~2011dh has a velocity of $v_{\rm LSR}\approx474$~km~s$^{-1}$ (component 7), 
which is close to the adopted recessional velocity of the galaxy of 
468~km~s$^{-1}$. At the apparent position of SN~2011dh, however, the velocity 
exhibited by disk gas should be significantly higher than the systemic 
velocity due to the inclination of M51 ($i=42^{\circ}$; Tamburro et al. 2008). 
The velocity field (moment 1 map) provided by The H~{\sc i} Nearby Galaxy 
Survey (THINGS; Walter et al. 2008) indicates a mean H~{\sc i} velocity of 
$v_{\rm LSR}\approx524$~km~s$^{-1}$ at the supernova position. The 
corresponding H~{\sc i} emission profile exhibits a single narrow 
(FWHM~$\approx$~18~km~s$^{-1}$) peak at this velocity, with no prominent 
emission at the velocity of component 7 (see Figure~3). In fact, none of the 
prominent Na~{\sc i} or Ca~{\sc ii} components in our spectra appear to 
coincide with the peak in H~{\sc i} emission, suggesting that the supernova 
lies just in front of the bulk of the H~{\sc i} disk.

Adopting the mean H~{\sc i} velocity from THINGS as the local disk velocity at 
the position of the supernova, the observed absorption from M51 ranges from 
$v-v_{\rm disk}\approx-79$~km~s$^{-1}$ (component 6) to $+63$~km~s$^{-1}$ 
(component 13). The velocity dispersion in Na~{\sc i} is $\sim$39~km~s$^{-1}$, 
and $\sim$33~km~s$^{-1}$ in Ca~{\sc ii}. Both are larger than the H~{\sc i} 
velocity dispersion, which is $\sim$20~km~s$^{-1}$ at the supernova position 
(based on the THINGS moment 2 map). The fact that we are detecting more 
components in Na~{\sc i} and Ca~{\sc ii} than are detected via H~{\sc i} 21~cm 
emission --- and detecting them over a broader range in velocity --- is most 
likely related to the limited sensitivity of the H~{\sc i} observations. The 
column density detection limit in THINGS channel maps of M51 is 
log~$N$(H~{\sc i})~$\sim20.1$ (for a 3$\sigma$ detection in at least two 
channels; see Walter et al. 2008). For comparison, the Na~{\sc i} column 
densities of components 6--13 imply H~{\sc i} column densities in the range 
log~$N$(H~{\sc i})~$\approx19.8$--$20.2$, indicating that much of this 
H~{\sc i} would be undetectable.\footnote{These estimates are based on the 
roughly quadratic relationship between $N$(Na~{\sc i}) and $N_{\rm tot}$(H) 
discussed by Welty \& Hobbs (2001), and the assumption that virtually all of 
the hydrogen is in neutral atomic rather than molecular form, as also implied 
by the Na~{\sc i} column densities. By applying these relationships, we 
implicitly assume that the metallicity and radiation field in the M51 
foreground are similar to typical Galactic values.} In contrast, the H~{\sc i} 
emission actually detected by THINGS at the position of SN~2011dh yields 
log~$N$(H~{\sc i})~=~$20.6\pm0.1$. If the supernova were positioned behind 
this material, we would expect to see Na~{\sc i} absorption with a column 
density approaching log~$N$(Na~{\sc i})~$\sim12.3$.

Still, the $N$(Na~{\sc i})/$N$(Ca~{\sc ii}) ratios exhibited by the M51 
components are similar to those seen in disk clouds of the Milky Way (e.g., 
Sembach \& Danks 1994) and in the disks of other external galaxies that have 
been probed by bright supernovae (e.g., Bowen et al. 2000). The components 
with higher Na~{\sc i} column densities generally have higher 
$N$(Na~{\sc i})/$N$(Ca~{\sc ii}) ratios, and at least two components have 
$N$(Na~{\sc i})/$N$(Ca~{\sc ii})~$\gtrsim1$, an indication that the absorption 
originates in relatively cool gas (see, e.g., Bertin et al. 1993). 
Interestingly, we find similar ratios (of order unity) in both quiescent gas, 
with $\vert v-v_{\rm disk}\vert<20$~km~s$^{-1}$, and higher-velocity clouds, 
with $\vert v-v_{\rm disk}\vert\gtrsim50$~km~s$^{-1}$, contrary to 
expectations based on the Routly-Spitzer effect (Routly \& Spitzer 1952). 
Thus, rather than showing evidence for enhanced grain destruction in 
high-velocity material, our results indicate that a similar degree of Ca 
depletion characterizes both quiescent and high-velocity gas in the portion of 
M51 sampled by this particular line of sight. The higher-velocity material may 
not originate very far from the disk, however, since the components have 
$\vert v-v_{\rm disk}\vert<100$~km~s$^{-1}$. These velocities are similar to 
those of intermediate-velocity clouds (IVCs) in the Milky Way, which are 
usually found within $\sim$5~kpc of the Galactic plane (e.g., 
Wakker et al. 2008).

Typical scenarios invoked to explain the origin of Galactic HVCs and IVCs 
include production in a Galactic fountain, tidal stripping from a companion 
galaxy, and accretion from the intergalactic medium, among others (see 
Wakker \& van Woerden 1997; Richter 2006). Since most of the Galactic IVCs 
have solar metalicities (Wakker 2001; Richter et al. 2001), their origin can 
most easily be explained within the context of the Galactic fountain model, in 
which hot gas is injected into the halo by supernova explosions and then cools 
and falls back onto the disk. A similar process may be responsible for the 
relatively cool cloud we find in the direction of SN~2011dh with a high 
positive velocity, which indicates that it is falling toward the disk of M51. 
On the other hand, tidal interactions between M51 and its companion (NGC 5195) 
may have played a role in generating the high velocity dispersion we observe 
in Na~{\sc i} and Ca~{\sc ii}. Tidal effects may be particularly important for 
explaining the origin of Na~{\sc i} components with high negative velocities, 
since in the outflow stage of the galactic fountain process the gas is 
expected to be mostly ionized (Fraternali et al. 2004). Multiple processes may 
be occurring simultaneously. Indeed, Miller \& Bregman (2005) invoke both 
galactic fountain and tidal stripping scenarios to explain deep Very Large 
Array observations of M51, which reveal a number of discrete H~{\sc i} sources 
with anomalous velocities.

An interesting result of Ho \& Filippenko's (1995) study of SN 1994I was their 
detection of Na~{\sc i} components at velocities intermediate between those 
expected for gas associated with the Milky Way and M51 (i.e., 
$v_{\rm LSR}\approx255$--$310$ km~s$^{-1}$; their systems 3--5). The authors 
attributed this absorption to HVCs, but could not distinguish between an 
origin in the Galactic halo or in the halo of M51, since the velocities of the 
clouds with respect to the overall systemic velocity of the galaxy to which 
they belong would be similar in either case. We detect no Na~{\sc i} or 
Ca~{\sc ii} absorption features in this velocity range toward SN~2011dh 
(just $2^{\prime}.3$ away), indicating that if the clouds indeed have a 
Galactic origin, they exhibit significant structure on small scales.

\acknowledgments
We thank Suzanne Hawley for granting us Director's Discretionary time at APO 
so that prompt spectra of SN 2011dh could be obtained and John Wisniewski for 
contributing the spectrum from June 13. We are also grateful for the comments 
and suggestions provided by the anonymous referee. This research was supported 
by the Kennilworth Fund of the New York Community Trust.

\clearpage

\begin{deluxetable}{ccccc}
\tablecolumns{5}
\tablewidth{0pt}
\tabletypesize{\scriptsize}
\tablecaption{Na~{\sc i} and Ca~{\sc ii} Equivalent Widths}
\tablehead{ \colhead{Comp.} & \colhead{$W_{\lambda}$(D$_2$)} & 
\colhead{$W_{\lambda}$(D$_1$)} & \colhead{$W_{\lambda}$(K)} & 
\colhead{$W_{\lambda}$(H)} \\
\colhead{No.} & \colhead{(m\AA)} & \colhead{(m\AA)} & \colhead{(m\AA)} & 
\colhead{(m\AA)} }
\startdata
\multicolumn{5}{c}{Milky Way} \\
\hline
1 & \ldots & \ldots & $16.3\pm3.4$ & $\phn9.4\pm3.7$ \\
2 & \ldots & \ldots & $30.7\pm3.4$ & $18.9\pm3.7$ \\
3 & $136.4\pm1.8$ & $113.7\pm1.8$ & $37.9\pm3.4$ & $23.3\pm3.7$ \\
4 & \ldots & \ldots & $22.4\pm3.0$ & $13.6\pm3.3$ \\
5 & \ldots & \ldots & $25.0\pm3.0$ & $15.7\pm3.3$ \\
\hline
\multicolumn{5}{c}{M51} \\
\hline
6 & $13.4\pm1.8$ & $\phn8.1\pm1.8$ & \ldots & \ldots \\
7 & $50.0\pm1.8$ & $32.1\pm1.8$ & $30.4\pm3.0$ & $17.0\pm3.3$ \\
8 & $12.5\pm1.7$ & $\phn7.5\pm1.7$ & $15.8\pm3.4$ & $\phn8.3\pm3.7$ \\
9 & $\phn6.8\pm1.7$ & $\phn4.7\pm1.7$ & $20.5\pm3.4$ & $11.0\pm3.7$ \\
10 & $30.9\pm1.8$ & $17.3\pm1.8$ & $13.1\pm3.4$ & $\phn7.0\pm3.7$ \\
11 & $\phn7.3\pm1.8$ & $\phn2.9\pm1.7$ & $10.1\pm2.9$ & $\phn5.3\pm3.2$ \\
12 & $40.0\pm1.8$ & $20.1\pm1.8$ & $20.4\pm2.9$ & $11.8\pm3.2$ \\
13 & $19.2\pm1.7$ & $13.4\pm1.8$ & $\phn9.4\pm3.0$ & $\phn4.8\pm3.3$ \\
\enddata
\end{deluxetable}

\begin{deluxetable}{ccccccccccc}
\setlength{\tabcolsep}{0.05in}
\tablecolumns{11}
\tablewidth{0pt}
\tabletypesize{\scriptsize}
\tablecaption{Column Densities and Na~{\sc i}/Ca~{\sc ii} Ratios}
\tablehead{ \colhead{} & \colhead{} & \multicolumn{2}{c}{Na~{\sc i}~D$_2$} & 
\colhead{} & \multicolumn{2}{c}{Na~{\sc i}~D$_1$} & \colhead{} & 
\multicolumn{2}{c}{Ca~{\sc ii}~K} & \colhead{} \\
\cline{3-4} \cline{6-7} \cline{9-10} \\
\colhead{Comp.} & \colhead{$v_{\mathrm{LSR}}$\tablenotemark{a}} & 
\colhead{$b$} & \colhead{log $N$} & \colhead{} & \colhead{$b$} & 
\colhead{log $N$} & \colhead{log $N$(Na~{\sc i})\tablenotemark{b}} & 
\colhead{$b$} & \colhead{log $N$} & \colhead{$N$(Na~{\sc i})/$N$(Ca~{\sc ii})} 
\\
\colhead{No.} & \colhead{(km s$^{-1}$)} & \colhead{(km s$^{-1}$)} & \colhead{} 
& \colhead{} & \colhead{(km s$^{-1}$)} & \colhead{} & \colhead{} & 
\colhead{(km s$^{-1}$)} & \colhead{} & \colhead{} }
\startdata
\multicolumn{11}{c}{Milky Way} \\
\hline
1 & $\phn-$$61.9\pm1.5$ & \ldots & \ldots && \ldots & \ldots & $<10.4$ & 3.8 & 
$11.31\pm0.08$ & $<0.1$ \\
2 & $\phn-$$49.0\pm1.5$ & \ldots & \ldots && \ldots & \ldots & $<10.4$ & 3.8 & 
$11.61\pm0.04$ & $<0.1$ \\
3 & $\phn-$$29.6\pm0.2$ & 2.1 & $12.48\pm0.14$ && 2.1 & $12.48\pm0.07$ & 
$12.48\pm0.06$ & 3.8 & $11.72\pm0.04$ & $5.7\pm1.0$ \\
4 & $\phn-$$12.9\pm1.5$ & \ldots & \ldots && \ldots & \ldots & $<10.4$ & 2.1 & 
$11.50\pm0.06$ & $<0.1$ \\
5 & $\phn\phn-$$0.4\pm1.5$ & \ldots & \ldots && \ldots & \ldots & $<10.4$ & 
2.1 & $11.55\pm0.05$ & $<0.1$ \\
\hline
\multicolumn{11}{c}{M51} \\
\hline
6 & $+445.2\pm1.5$ & 2.6 & $10.85\pm0.06$ && 2.6 & $10.93\pm0.09$ & 
$10.86\pm0.05$ & \ldots & $<11.0$ & $>0.7$ \\
7 & $+473.7\pm1.0$ & 2.1 & $11.52\pm0.02$ && 2.6 & $11.57\pm0.02$ & 
$11.53\pm0.01$ & 2.1 & $11.66\pm0.04$ & $0.7\pm0.1$ \\
8 & $+482.0\pm0.3$ & 0.5 & $10.92\pm0.06$ && 1.3 & $10.90\pm0.09$ & 
$10.92\pm0.05$ & 3.8 & $11.29\pm0.08$ & $0.4\pm0.1$ \\
9 & $+498.0\pm1.5$ & 0.6 & $10.58\pm0.10$ && 0.5 & $10.72\pm0.13$ & 
$10.61\pm0.08$ & 3.8 & $11.42\pm0.07$ & $0.2\pm0.1$ \\
10 & $+508.8\pm2.8$ & 2.4 & $11.25\pm0.02$ && 2.6 & $11.27\pm0.04$ & 
$11.25\pm0.02$ & 3.8 & $11.21\pm0.10$ & $1.1\pm0.3$ \\
11 & $+523.2\pm5.0$ & 2.1 & $10.58\pm0.10$ && 1.4 & $10.48\pm0.20$ & 
$10.56\pm0.09$ & 1.2 & $11.13\pm0.11$ & $0.3\pm0.1$ \\
12 & $+540.3\pm0.2$ & 2.6 & $11.37\pm0.02$ && 2.6 & $11.34\pm0.04$ & 
$11.37\pm0.02$ & 1.0 & $11.54\pm0.06$ & $0.7\pm0.1$ \\
13 & $+587.4\pm0.8$ & 1.4 & $11.05\pm0.04$ && 1.9 & $11.16\pm0.05$ & 
$11.08\pm0.03$ & 2.1 & $11.07\pm0.12$ & $1.0\pm0.3$ \\
\enddata
\tablecomments{Upper limits on column densities for nondetections are 
3$\sigma$.}
\tablenotetext{a}{Mean (and 1$\sigma$ standard deviation) of the velocities 
found in Na~{\sc i}~D$_2$ and Ca~{\sc ii}~K.}
\tablenotetext{b}{Weighted mean of the column densities derived from the 
Na~{\sc i}~D$_2$ and D$_1$ lines.}
\end{deluxetable}

\end{document}